\documentclass[10pt,floatfix]{revtex4}
\usepackage{bm}
\usepackage{amsmath}
\usepackage{graphics}

\newcommand{\vc}[1]{\ensuremath{\bm{#1}}}
\newcommand{\vh}[1]{\ensuremath{\hat{\bm{#1}}}}
\newcommand{\avg}[1]{\ensuremath{\langle #1 \rangle}}
\newcommand{\ThreeJ}[6]{\left (\begin{array}{ccc}#1&#2&#3\\
#4&#5&#6\end{array} \right )}
\newcommand{\SixJ}[6]{\left \{\begin{array}{ccc}#1&#2&#3\\
#4&#5&#6\end{array} \right \}}
\newcommand{\SN}{\left (\frac{S}{N}\right )^2}
\newcommand{\xf}{x}
\newcommand{\yf}{y}
\newcommand{\zf}{z}
\newcommand{\wf}{w}
\newcommand{\af}{a}
\newcommand{\bff}{b}
\newcommand{\cf}{c}
\newcommand{\df}{d}
\newcommand{\ind}[2]{#1#2}
\newcommand{\rarg}[3]{#1 #2 #3}
\newcommand{\narg}[2]{#1 #2}
\newcommand{\myfigure}[1]{\resizebox{3in}{!}{\includegraphics{#1}}}

\def\edth{\;\raise1.0pt\hbox{$'$}\hskip-6pt\partial\;}
\def\baredth{\;\overline{\raise1.0pt\hbox{$'$}\hskip-6pt
\partial}\;}

\begin{document}
\title{The Angular Trispectra of CMB Temperature and Polarization}
\date{\today}
\author{Takemi Okamoto}
\affiliation{Department of Physics, University of Chicago, Chicago, IL 60637}
\email{tokamoto@oddjob.uchicago.edu}
\author{Wayne Hu}
\affiliation{Center for Cosmological Physics and the
Department of Astronomy and Astrophysics and the Enrico Fermi Institute, 
University of Chicago, Chicago, IL 60637}
\email{whu@background.uchicago.edu}
\begin{abstract}
We develop the formalism necessary to study four-point functions of the
cosmic 
microwave background (CMB) temperature and polarization fields.  We determine the 
general form of CMB trispectra, with the constraints imposed by the 
assumption of statistical isotropy of the CMB fields, and derive expressions 
for their estimators, as well as their Gaussian noise properties.  We apply
these techniques to initial non-Gaussianity of a form motivated by inflationary
models.  Due to the large number of four-point configurations, the sensitivity
of the trispectra to initial non-Gaussianity approaches that of the 
temperature bispectrum at high multipole moment.  These trispectra techniques will also be useful in
the study of secondary anisotropies induced for example by the gravitational
lensing of the CMB by the large scale structure of the universe.
\end{abstract}
\maketitle

\section{Introduction\label{Sect:Introduction}}

Beyond the power spectra of the cosmic microwave background (CMB)
lies the relatively unexplored territory of non-Gaussian statistics.  Studies of its
non-Gaussianity hold the potential to reveal physics at the two 
ends of time.  Non-Gaussianity in the primary anisotropies from
the recombination epoch can test the inflationary model of the origin
of fluctuations (e.g., \cite{SandB1,Gangui1,Gangui2,Falk1,Munshi1}).  
Non-Gaussianity in the secondary anisotropies, arising during
the transit of a CMB photon through the large-scale
structure of the universe, probes the nature of the dark energy and dark matter
(e.g., \cite{Bernardeau1,PandC,Spergel2,Zaldarriaga1,HuDE}).

The primary challenge facing non-Gaussian studies of the CMB is the selection of 
an appropriate statistics.  The term ``non-Gaussianity'' tells us what the
distribution is not, not what it is. Like the power spectra,
the higher-point correlations of the multipole moments of CMB fields 
provide a set of statistics with definitive predictions in the cases of 
cosmological interest.  Unlike the power spectra, there are a large
number of potential observables, associated with the distinct configurations
of the points, requiring the development of new techniques for
their prediction and estimation.  In particular, it is important to
identify the symmetry properties of the spectra to build optimal
statistics for the detection of non-Gaussianity.

Non-Gaussian signatures in the three-point function or bispectrum of the
temperature distribution
\cite{Gangui1,Spergel2,Verde1, Cooray1,Komatsu1} and
polarization \cite{Hu2} as well as techniques for their
extraction \cite{Luo1,Heavens1,Spergel1,Sandvik1,Phillips1,Komatsu3}
have been studied extensively in the past few years.
The four-point function or trispectrum has recently 
received much attention due to its use in the study of the
gravitational lensing of the CMB 
\cite{Bernardeau1,Zaldarriaga1,HuTrispec,Cooray2}.
Techniques for measuring certain components have been tested
on the Cosmic Background Explorer (COBE) data \cite{KomatsuPhD,Kunetal01}.
Still, a complete treatment incorporating the full symmetry properties
of the temperature and polarization fields has been lacking in the 
literature.

In this paper we complete the formalism established for the
temperature trispectrum \cite{HuTrispec}.  The addition of polarization 
information leads to a multiplicity of trispectra corresponding to 
all possible combinations of three observable fields.   It has been recently shown that
the higher point correlations of the CMB polarization contain the majority of
the information on gravitational lensing in the CMB \cite{HuOka02}. 
Trispectra also quantify the non-Gaussian errors to temperature and polarization
power spectra measurements.

The outline of the paper is as follows. 
We consider the general
symmetry and noise properties of trispectra in Sec.~\ref{Sect:Formalism}.
As an illustration of the construction and noise properties of trispectra,
we apply these techniques to the initial non-Gaussianity in the curvature
fluctuations of the form predicted by slow-roll inflation in Sec.~\ref{Sect:Initial}.
We show that the sensitivity
to initial non-Gaussianity in the trispectra can approach that 
in the temperature bispectrum \cite{Komatsu1}.
In Appendix \ref{Appendix:SpinS} we summarize relations useful
for the study of high order correlations in the polarization.  In Appendix 
\ref{Appendix:Additional} we cover the details in the properties, measurement,
 and approximation of the trispectra that may be useful for future studies.

\section{Formalism\label{Sect:Formalism}}

We begin with definitions associated with the harmonic description of
the temperature and polarization fields in Sec.~\ref{Sect:Definitions}.
We consider the general symmetry properties of $n$-point harmonic functions 
in Sec.~\ref{Sect:Invariance} and apply them to the trispectra
in Sec.~\ref{Sect:Trispectra}.  Finally we derive the Gaussian noise properties
of trispectra estimators in Sec.~\ref{Sect:Noise}.

\subsection{Definitions}
\label{Sect:Definitions}

The temperature field $\Theta(\vh{n})\equiv\Delta T(\vh{n})/T$
is decomposed into multipole moments according to
\begin{eqnarray}
\Theta(\vh{n})& = &\sum_{lm}\Theta_l^m Y_l^m(\vh{n}).
\label{ThetaDecompose}
\end{eqnarray}
The polarization, described by the Stokes
parameters $Q(\vh{n})$ and $U(\vh{n})$ in spherical polar coordinates,
is a spin-2 field and is
similarly decomposed as \cite{ZandS1,Kam1}
\begin{eqnarray}
(Q\pm iU)(\vh{n}) &=& \sum_{lm}{}_\pm A_l^m {}_{\pm 2}Y_l^m(\vh{n}),
\label{A_Coeffs}
\end{eqnarray}
where ${}_{s}Y_l^m(\vh{n})$ is the spin-weighted spherical harmonics
whose properties are reviewed  
in Appendix~\ref{Appendix:SpinS}.  Note that ${}_0 Y_l^m = Y_l^m$.

Under a parity transformation $\hat{P}$ taking $\vh{n}$ to $-\vh{n}$, the 
spin-spherical harmonics transform as 
\begin{eqnarray}
\hat{P}[{}_sY_l^m(\vh{n})]&=& (-1)^l{}_{-s}Y_l^m(\vh{n}),
\label{SpinSParity}
\end{eqnarray}
and so it is convenient to define the parity eigenfunctions
\begin{eqnarray}
{}_s\mathcal{E}_l^m&\equiv&\frac{{}_sY_l^m+{}_{-s}Y_l^m}{2},\\
{}_s\mathcal{O}_l^m&\equiv&\frac{{}_sY_l^m-{}_{-s}Y_l^m}{2i}.
\end{eqnarray}
We see that under parity, 
$\hat{P}[{}_s\mathcal{E}_l^m]=(-1)^l{}_s\mathcal{E}_l^m$,
whereas $\hat{P}[{}_s\mathcal{O}_l^m]=(-1)^{l+1}{}_s\mathcal{O}_l^m$.  
A spin-0 field such as the temperature fluctuation 
carries only the ${}_0 \mathcal{E}_l^m = Y_l^m$ mode.  The spin-2 polarization
field has two components that are distinguished by parity \cite{Kam1,ZandS1} 
\begin{eqnarray}
{}_\pm A_l^m &=& E_l^m \pm iB_l^m
\label{ParityStates}
\end{eqnarray}
called the $E$ and the $B$ modes.
This definition differs from that in \cite{ZandS1} by an overall minus
sign, so in particular, the sign of temperature-polarization cross-correlations is
reversed.  

The fields $E(\vh{n})$ and $B(\vh{n})$ on the sky are defined as 
\begin{subequations}
\label{RealEBFields}
\begin{eqnarray}
E(\vh{n})&=&\sum_{lm} E_l^m Y_l^m(\vh{n})\label{RealEField},\\
B(\vh{n})&=&\sum_{lm} B_l^m Y_l^m(\vh{n})\label{RealBField}.
\end{eqnarray}
\end{subequations}
The parity properties of the eigenstates ${}_s\mathcal{E}_l^m$ and 
${}_s\mathcal{O}_l^m$ imply that $E(\vh{n})$ will be a scalar under parity, 
whereas $B(\vh{n})$ will be a pseudoscalar.

Lastly, requiring that the fields $\Theta(\vh{n})$, $Q(\vh{n})$, and 
$U(\vh{n})$ be real imposes the constraints
\begin{eqnarray}
\xf_l^{m*} &= &(-1)^m\xf_l^{-m}, \xf\in\{\Theta,E,B\}
\label{RealityCondition}
\end{eqnarray}
on the multipole moments.  This constraint also enforces the reality of
$E(\vh{n})$ and $B(\vh{n})$.

\subsection{Rotational and parity invariance}
\label{Sect:Invariance}

The harmonic transform of the $n$-point correlation of CMB fields 
$\xf, \ldots, \zf \in\{\Theta,E,B\}$ defines the $n$-point harmonic correlation 
functions according to
\begin{eqnarray}
\avg{\xf(\vh{n}_1)\ldots \zf(\vh{n}_n)}&=&\sum_{l_i,m_i}
\avg{\xf_{l_1}^{m_1}\ldots \zf_{l_n}^{m_n}}Y_{l_1}^{m_1}(\vh{n}_1)
\ldots Y_{l_n}^{m_n}(\vh{n}_n).
\end{eqnarray}
Since the CMB fields are assumed to be  statistically
isotropic, the $n$-point correlations must be 
invariant under rotations.  A general rotation
 $\vh{R}$ acts on spherical harmonics as
\begin{eqnarray}
\vh{R}[Y_l^m(\vh{n})]& =& \sum_{m'}D^l_{\ind{m}{m'}}(\vh{R})Y_l^{m'}(\vh{n}).
\end{eqnarray}
Useful properties of the rotation matrix $D^l_{\ind{m}{m'}}$ are summarized
in Appendix \ref{Appendix:SpinS}.
The $n$-point harmonic correlation must
therefore obey
\begin{eqnarray}
\langle \xf_{l_1}^{m_1}\ldots \zf_{l_n}^{m_n} \rangle&=&
\sum_{m'_1\ldots m'_n}
\langle \xf_{l_1}^{m'_1}\ldots \zf_{l_n}^{m'_n} \rangle
D^{l_1}_{\ind{m_1}{m'_1}}(\vh{R})\ldots D^{l_n}_{\ind{m_n}{m'_n}}(\vh{R}).
\label{nPointRotation}
\end{eqnarray}
This invariance demands a specific form for the $m$-dependence as we shall see.

Under a parity transformation
an $n$-point function
containing $k$ $B$ fields will transform according to
\begin{eqnarray}
&&\hat{P}[\avg{\xf_{l_1}^{m_1}\ldots \zf_{l_n}^{m_n}}]
= (-1)^{k+\sum l_i} \avg{\xf_{l_1}^{m_1}\ldots \zf_{l_n}^{m_n}}.
\label{ParityTransform}
\end{eqnarray}
Invariance under a parity transformation therefore implies that the 
$n$-point function containing $k$ $B$ fields will vanish when
\begin{eqnarray}
k+\sum_i l_i =\text{ odd}.
\label{ParityCondition}
\end{eqnarray}

\subsection{Trispectra}
\label{Sect:Trispectra}

\begin{figure}[tbhp]
\myfigure{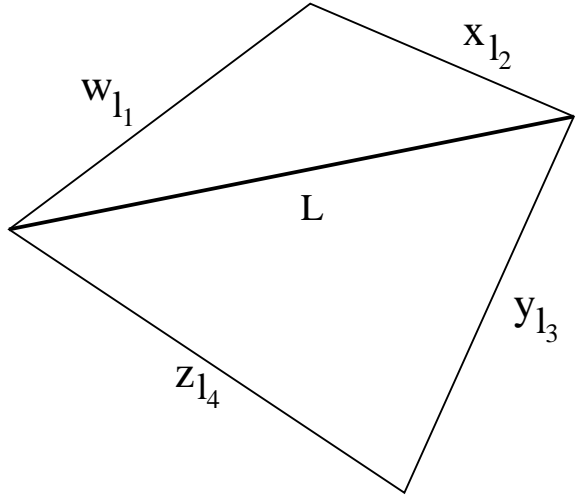}
\caption{Geometrical interpretation of the configuration of a trispectrum.
The four-point quadrilateral in harmonic space is specified using the 
pairs $(l_1,l_2)$ along with the diagonal $L$ to define a triangle.}
\label{fig1}
\end{figure}

The reduction of the four-point harmonic function into a rotationally
invariant form  follows the steps outlined 
in \cite{HuTrispec}.  Using the Clebsch-Gordan property 
(\ref{DFunctionMult}) on 
Eq. (\ref{nPointRotation}) to pair  
$(l_1,l_2)$ and $(l_3,l_4)$ together and applying the orthogonality
condition (\ref{AppEqn:DOrtho}) to the resultant pair,
we reduce the function to
\begin{eqnarray}
\avg{\wf_{l_1}^{m_1}\xf_{l_2}^{m_2}\yf_{l_3}^{m_3}\zf_{l_4}^{m_4}}
&\equiv&\sum_{LM}(-1)^M\ThreeJ{l_1}{l_2}{L}{m_1}{m_2}{-M}
\ThreeJ{l_3}{l_4}{L}{m_3}{m_4}{M}
Q^{\wf_{l_1}\xf_{l_2}}_{\yf_{l_3}\zf_{l_4}}(L).
\label{TrispecDefinition}
\end{eqnarray}
The trispectrum $Q^{\wf_{l_1}\xf_{l_2}}_{\yf_{l_3}\zf_{l_4}}(L)$ represents a 
configuration with sides $l_1\ldots l_4$ labeled by the fields $\wf\ldots \zf$,
with one diagonal of length $L$ forming a triangle with $l_1$ and $l_2$
(Fig.~\ref{fig1}).

Choosing the two other pairings $(l_1,l_3)$ and $(l_1,l_4)$ yield alternate
representations of the trispectra.  Since each representation
is constructed by adding pairs of angular momenta, each representation is 
complete, and three representations are related through Wigner 6-$j$
symbols via
\begin{subequations}
\label{Recoupling}
\begin{eqnarray}
Q^{\wf_{l_1}\yf_{l_3}}_{\xf_{l_2}\zf_{l_4}}(L)&=&\sum_{L'}(-1)^{l_2+l_3}(2L+1)
\SixJ{l_1}{l_2}{L'}{l_4}{l_3}{L}
Q^{\wf_{l_1}\xf_{l_2}}_{\yf_{l_3}\zf_{l_4}}(L')
\label{Recouple1}
\end{eqnarray}
and 
\begin{eqnarray}
Q^{\wf_{l_1}\zf_{l_4}}_{\yf_{l_3}\xf_{l_2}}(L)&=&\sum_{L'}(-1)^{L+L'}(2L+1)
\SixJ{l_1}{l_2}{L'}{l_3}{l_4}{L}
Q^{\wf_{l_1}\xf_{l_2}}_{\yf_{l_3}\zf_{l_4}}(L').
\label{Recouple2}
\end{eqnarray}
\end{subequations}

The trispectrum is obtained by subtracting the unconnected or Gaussian piece
from $Q^{\wf_{l_1}\xf_{l_2}}_{\yf_{l_3}\zf_{l_4}}(L)$, so
\begin{eqnarray}
T^{\wf_{l_1}\xf_{l_2}}_{\yf_{l_3}\zf_{l_4}}(L)&=&Q^{\wf_{l_1}\xf_{l_2}}_{\yf_{l_3}\zf_{l_4}}(L)
-G^{\wf_{l_1}\xf_{l_2}}_{\yf_{l_3}\zf_{l_4}}(L),
\end{eqnarray}
where the Gaussian piece is constructed from the power spectra 
\begin{eqnarray}
C_l^{xy} = \left< x_l^{m*} y_l^m \right>
\end{eqnarray}
as
\begin{eqnarray}
G^{\wf_{l_1}\xf_{l_2}}_{\yf_{l_3}\zf_{l_4}}(L)&=&
(-1)^{l_1+l_3}\sqrt{(2l_1+1)(2l_3+1)}
C_{l_1}^{\wf\xf}C_{l_3}^{\yf\zf}
\delta_{l_1l_2}\delta_{l_3l_4}\delta_{L0}\nonumber \\
&+&(2L+1)
\left [(-1)^{l_1+l_2+L}C_{l_1}^{\wf\yf}C_{l_2}^{\xf\zf}\delta_{l_1l_3}
\delta_{l_2l_4}+C_{l_1}^{\wf\zf}C_{l_2}^{\xf\yf}\delta_{l_1l_4}\delta_{l_2l_3}\right ].
\label{Gaussian}
\end{eqnarray}

From the permutation symmetry of the trispectrum (\ref{TrispecDefinition}),
additional constraints hold
\begin{subequations}
\label{TrispecConstraints}
\begin{eqnarray}
T^{\wf_{l_1}\xf_{l_2}}_{\yf_{l_3}\zf_{l_4}}(L)=
(-1)^{\Sigma_U}T^{\xf_{l_2}\wf_{l_1}}_{\yf_{l_3}\zf_{l_4}}(L)=
(-1)^{\Sigma_L}T^{\wf_{l_1}\xf_{l_2}}_{\zf_{l_4}\yf_{l_3}}(L)=
(-1)^{\Sigma_U+\Sigma_L}T^{\xf_{l_2}\wf_{l_1}}_{\zf_{l_4}\yf_{l_3}}(L),
\end{eqnarray}
where
\begin{eqnarray}
\Sigma_U &=&L+l_1+l_2,\\
\Sigma_L &=&L+l_3+l_4.
\end{eqnarray}
\end{subequations}
The constraints (\ref{Recoupling}) and (\ref{TrispecConstraints}) express 
redundancies in the definition of the trispectrum, where a physical
configuration can be labeled by $4!=24$ different permutations of the field
labels and pairings.  

In practice, the following construction guarantees that trispectra obey the 
constraints outlined above.  Given that the connected part of the four-point 
function can be expanded into its three pairings as
\begin{eqnarray}
\avg{\wf_{l_1}^{m_1}\xf_{l_2}^{m_2}\yf_{l_3}^{m_3}\zf_{l_4}^{m_4}}_c&=&
\sum_{LM}(-1)^M\ThreeJ{l_1}{l_2}{L}{m_1}{m_2}{-M}
\ThreeJ{l_3}{l_4}{L}{m_3}{m_4}{M}
P^{\wf_{l_1}\xf_{l_2}}_{\yf_{l_3}\zf_{l_4}}(L)\nonumber \\
&+&(\xf_{l_2}\leftrightarrow \yf_{l_3})
+(\xf_{l_2}\leftrightarrow \yf_{l_4}),
\label{FourPtToTrispec}
\end{eqnarray}
the latter two pairings are projected onto the $(l_1,l_2)$ basis through the
recoupling relations (\ref{Recoupling}) to give
\begin{eqnarray}
T^{\wf_{l_1}\xf_{l_2}}_{\yf_{l_3}\zf_{l_4}}(L)&=&
P^{\wf_{l_1}\xf_{l_2}}_{\yf_{l_3}\zf_{l_4}}(L)+(2L+1)\sum_{L'}
\left [(-1)^{l_2+l_3}\SixJ{l_1}{l_2}{L}{l_4}{l_3}{L'}
P^{\wf_{l_1}\yf_{l_3}}_{\xf_{l_2}\zf_{l_4}}(L')\right.\nonumber \\
&+&\left.(-1)^{L+L'}\SixJ{l_1}{l_2}{L}{l_3}{l_4}{L'}
P^{\wf_{l_1}\zf_{l_4}}_{\yf_{l_3}\xf_{l_2}}(L') \right ].
\label{TrispecEnforced1}
\end{eqnarray}
Having satisfied the recoupling relations, the remaining constraints 
(\ref{TrispecConstraints}) are enforced by introducing a reduced function 
$\mathcal{T}^{\wf_{l_1}\xf_{l_2}}_{\yf_{l_3}\zf_{l_4}}(L)$, where
\begin{eqnarray}
P^{\wf_{l_1}\xf_{l_2}}_{\yf_{l_3}\zf_{l_4}}(L)&=&
\mathcal{T}^{\wf_{l_1}\xf_{l_2}}_{\yf_{l_3}\zf_{l_4}}(L)+
(-1)^{\Sigma_U}\mathcal{T}^{\xf_{l_2}\wf_{l_1}}_{\yf_{l_3}\zf_{l_4}}(L)+
(-1)^{\Sigma_L}\mathcal{T}^{\wf_{l_1}\xf_{l_2}}_{\zf_{l_4}\yf_{l_3}}(L)+
(-1)^{\Sigma_U+\Sigma_L}\mathcal{T}^{\xf_{l_2}\wf_{l_1}}_{\zf_{l_4}\yf_{l_3}}(L),
\label{CurlyTFunction}
\end{eqnarray}
with the additional constraint
\begin{eqnarray}
\mathcal{T}^{\wf_{l_1}\xf_{l_2}}_{\yf_{l_3}\zf_{l_4}}(L)
=\mathcal{T}^{\yf_{l_3}\zf_{l_4}}_{\wf_{l_1}\xf_{l_2}}(L).
\end{eqnarray}
These exhaust the $4!$ redundancies due to permutations of the fields. 
Given the functional form of $\mathcal{T}$, the trispectrum can be constructed
by permuting the fields along with their indices, as indicated above.  
Therefore $\mathcal{T}$ provides the most economical description of
the trispectra for a given physical effect.

It is possible to think of the trispectra configurations as 
being labeled by a fixed field configuration [e.g., 
$T^{\xf_{l_1}\yf_{l_2}}_{\xf_{l_3}\yf_{l_4}}(L)$ with an $\xf$ and $\yf$ always related
through the diagonal $L$], with the indices $l_i$ allowed to vary.  
The derivation of symmetry 
properties in such a viewpoint is straightforward, and is deferred to 
Appendix~\ref{Appendix:FixedFields}.  The above construction automatically 
enforces the symmetries with fixed field configurations, Eqs.
(\ref{XXXXConstraints})-(\ref{XXYZConstraints}).

\subsection{Noise properties\label{NoiseProperties}}
\label{Sect:Noise}

By inverting definition (\ref{TrispecDefinition}), we obtain an estimator for 
a trispectrum
\begin{eqnarray}
\hat{T}^{\wf_{l_1}\xf_{l_2}}_{\yf_{l_3}\zf_{l_4}}(L)&=&\sum_{m_i,M}(2L+1)(-1)^M
\ThreeJ{l_1}{l_2}{L}{m_1}{m_2}{M}\ThreeJ{l_3}{l_4}{L}{m_3}{m_4}{-M}
\avg{\wf_{l_1}^{m_1}\xf_{l_2}^{m_2}\yf_{l_3}^{m_3}\zf_{l_4}^{m_4}}
-\hat{G}^{\wf_{l_1}\xf_{l_2}}_{\yf_{l_3}\zf_{l_4}}(L),
\label{TrispecEstimator}
\end{eqnarray}
where the Gaussian estimator $\hat{G}$ is constructed using expression 
(\ref{Gaussian}) with the power spectra replaced by their estimators.  
We discuss more practical forms of this estimator in Appendix \ref{Appendix:Measurement}.

The covariance between two trispectrum estimators due to Gaussian noise then
becomes
\begin{eqnarray}
\frac{\avg{\hat{T}^{\af_{l_1}\bff_{l_2}}_{\cf_{l_3}\df_{l_4}}{}^*(L)
\hat{T}^{\wf_{l_1'}\xf_{l_2'}}_{\yf_{l_3'}\zf_{l_4'}}(L')}}{2L+1}&=&
\delta_{LL'} N^{12}_{34} 
+(2L'+1) \left [(-1)^{l_2+l_3}\left \{\begin{array}{ccc}l_1&l_2&L\\l_4&l_3&L'\end{array}
 \right \}
N^{13}_{24}
+ (-1)^{L+L'}\left \{\begin{array}{ccc}l_1&l_2&L\\
l_3&l_4&L'\end{array}\right \}
N^{14}_{32}
\right ],
\label{GeneralCovariance}
\end{eqnarray}
where no two $l$'s in the primed and unprimed sets are equal, and 
\begin{eqnarray}
N^{i j}_{k l}&=&
\left [(-1)^{L'+l_1+l_2}\delta_{l_1l'_i}\delta_{l_2l'_j}C_{l_1}^{\af f_i}
C_{l_2}^{\bff f_j}+\delta_{l_1l'_j}\delta_{l_2l'_i}
C_{l_1}^{\af f_j}C_{l_2}^{\bff f_i}\right ] 
\left [(-1)^{L'+l_3+l_4}\delta_{l_3l'_k}\delta_{l_4l'_l}
C_{l_3}^{\cf f_k}C_{l_4}^{\df f_l}+\delta_{l_3l'_l}\delta_{l_4l'_k}
C_{l_3}^{\cf f_l}C_{l_4}^{\df f_k}
\right ] 
\nonumber \\
&& + [i \leftrightarrow k,
j \leftrightarrow l],
\label{GeneralDelta}
\end{eqnarray}
with $f_i$ denoting the field associated with $l_i'$.
If any two $l_i$'s are equal, one would need to consider additional terms in
the covariance arising from pairings within the primed and unprimed sets.

Using the above covariance, the total signal-to-noise ratio 
is then given by
\begin{eqnarray}
\SN &=&\sum_{l_i,l_i'}\sum_{L,L'}
\sum_{abcd}\sum_{wxyz}
\avg{\hat{T}^{a_{l_1}b_{l_2}}_{c_{l_3}d_{l_4}}(L)}[\text{Cov}]^{-1}
\avg{\hat{T}^{w_{l_1'}x_{l_2'}}_{y_{l_3'}z_{l_4'}}(L')}.
\label{GeneralSN}
\end{eqnarray}
``Cov'' here is the covariance in Eq. (\ref{GeneralCovariance}) viewed as a 
matrix and the field-type sums are over the measured fields. 

This matrix will possess singular values for permutations that are
equivalent. They can be eliminated by singular value decomposition or
equivalently by restricting the sums to a set of inequivalent 
permutations.  The latter is computationally more efficient 
and the redundancies expressed 
in Eqs. (\ref{Recoupling}) and (\ref{TrispecConstraints}) can be removed by 
restricting the $l$ sums.  Thus the total signal-to-noise ratio simplifies to 
\begin{eqnarray}
\SN =\sum_{l_1>l_2>l_3>l_4}\sum_L
\sum_{abcd}
\sum_{wxyz}
\avg{\hat{T}^{a_{l_1}b_{l_2}}_{c_{l_3}d_{l_4}}(L)}
[\text{Cov}]^{-1}
\avg{\hat{T}^{w_{l_1}x_{l_2}}_{y_{l_3}z_{l_4}}(L)},
\label{SNtemp}
\end{eqnarray}
where the covariance between $\hat{T}^{a_{l_1}b_{l_2}}_{c_{l_3}d_{l_4}}(L)$
and $\hat{T}^{w_{l_1}x_{l_2}}_{y_{l_3}z_{l_4}}(L)$ likewise simplifies to 
\begin{eqnarray}
[\text{Cov}] &=&(2L+1)C_{l_1}^{aw}C_{l_2}^{bx}C_{l_3}^{cy}C_{l_4}^{dz},
\label{SNtempCovariance}
\end{eqnarray}
so that the matrix inverse is now only over field choices for a fixed
set of multipoles.

\section{Trispectra from Initial Conditions}
\label{Sect:Initial}

As an application of the formalism for describing temperature and polarization
trispectra and computing the signal-to-noise ratio of their estimators, we consider
here the signature of non-Gaussianity that is inherent in the initial conditions.  
In Secs. \ref{Sect:InflationNonLin} and \ref{Sect:InflationTrispec} 
we motivate a  form for the trispectra
based on slow-roll inflation.  Although typical models predict amplitudes far
below that which is potentially observable,  this form is
generic to local non-Gaussianity in the initial conditions.  In Sec. 
\ref{Sect:TPInfl}
we describe the transfer of this initial non-Gaussianity to the observable
temperature and polarization fields.  We show in Sec. \ref{Sect:SNInfl} that 
the
total signal-to-noise ratio in the trispectra is comparable to that in the 
temperature bispectrum considered previously in the literature.

\subsection{Inflationary motivation}
\label{Sect:InflationNonLin}

The standard inflationary paradigm is known
to predict a very nearly Gaussian spectrum of initial curvature fluctuations
which under linear gravitational instability theory implies a  Gaussian spectrum 
of the CMB fluctuations.  
However, nonlinear corrections in inflation and gravity can produce non-Gaussian 
fluctuations, which may be observable in 
the microwave background.  The imprint of such nonlinearity on the bispectrum
of the microwave background has been studied extensively.  The theoretical
predictions for the bispectrum and the related statistic of skewness has been
described in~\cite{Komatsu1,Gangui1,Gangui2,Phillips1}, and observational 
limits placed using existing data in~\cite{Komatsu2,Sandvik1,Verde1}.  We here
extend these considerations to higher order to investigate effects on the
trispectra.

Following~\cite{SandB1}, let us consider the non-Gaussianity induced in
corrections to the correspondence between a Gaussian inflaton fluctuation
and the Newtonian curvature $\Phi$.   $\Phi$ during matter domination can be related to the 
inflaton fluctuation at horizon exit according to
\begin{eqnarray}
\Phi(\vc{x})&=&\frac{12\pi G}{5}\int_{\phi_0}^{\phi_0+\delta\phi}\left [
\frac{\partial\ln H}{\partial\phi} \right ]^{-1}d\phi\nonumber \\
&\approx & \frac{12\pi G}{5}\left [\frac{\partial\ln H}{\partial\phi} 
\right ]^{-1}\delta\phi +\frac{6\pi G}{5}\frac{\partial}{\partial\phi}
\left [\frac{\partial\ln H}{\partial\phi} \right ]^{-1}\delta\phi^2
+\frac{2\pi G}{5}\frac{\partial^2}{\partial\phi^2}
\left [\frac{\partial\ln H}{\partial\phi} 
\right ]^{-1}\delta\phi^3+\mathcal{O}(\delta\phi^4).
\label{DeltaphiToPhi}
\end{eqnarray}
The leading order term, which we will denote
$\Phi_L(\vc{x})$, carries Gaussian random fluctuations from $\delta\phi$.
The higher order terms can then be written as
\begin{eqnarray}
\Phi(\vc{x})&=&\Phi_L(\vc{x})+f_1\left (\Phi_L^2(\vc{x})-
\avg{\Phi_l^2(\vc{x})} \right )+f_2\Phi_L^3(\vc{x})+
\mathcal{O}(\Phi_L^4),
\label{PhiNLCouple}
\end{eqnarray}
with
\begin{eqnarray}
f_1&=&-\frac{5}{6} \frac{1}{8\pi G}\frac{\partial^2\ln V}{\partial\phi^2},\nonumber \\
f_2&=&\frac{25}{54}\frac{1}{(8\pi G)^2}\left [2\left (
\frac{\partial^2\ln V}{\partial\phi^2} \right )^2-
\frac{\partial^3\ln V}{\partial\phi^3}
\frac{\partial\ln V}{\partial\phi}\right ]\,,
\end{eqnarray}
where $V(\phi)$ is the inflaton potential.
This model generalizes the considerations of
\cite{Coles1,Moscardini1,Falk1,Verde1,Komatsu1} to higher order.
Note that our $f_1$ corresponds to $f_{NL}$ in~\cite{Komatsu1}.
As an example consider an inflaton potential of the
form $V=\lambda \phi^n$, then
\begin{eqnarray}
f_1&=&\frac{5n}{6}\left (\frac{M_p}{\phi}\right )^2,\\
f_2&=& 0,
\end{eqnarray}
with $M_p^2\equiv 1/8\pi G$ defining the reduced Planck mass.  For inflation to
occur, this class of models requires that $\phi\approx\sqrt{120n}M_p$, so that
we obtain $f_1\sim 0.01$ 
as an order-of-magnitude estimate.  

Although these small coupling coefficients and the observed $10^{-5}$ level
of curvature perturbations make  non-Gaussian contributions from typical 
inflationary models unmeasurable,  the general form in Eq.~(\ref{PhiNLCouple}) 
is simply a perturbative expansion of the curvature
fluctuations 
and so we leave $f_1$ and $f_2$ as free
parameters and explore the extent to which they are measurable in
temperature and polarization trispectra.
Other sources of non-Gaussianity of this form include
 interaction terms in the inflaton potential~\cite{Falk1},
 stochastic interactions of the long-wavelength inflaton fluctuations
with the short-wavelength modes~\cite{SandB2,Gangui2}, and some multiple 
field models~\cite{Bartolo1}.

The ansatz (\ref{PhiNLCouple}) for the curvature fluctuations imply
higher order correlations in Fourier space since products become convolutions.
We can decompose the contributions into linear 
and nonlinear parts, so that
\begin{eqnarray}
\Phi(\vc{k})&\equiv&\Phi_L(\vc{k})+\Phi_A(\vc{k})+\Phi_B(\vc{k}),
\label{PhiNLKspace}
\end{eqnarray}
with
\begin{eqnarray}
\Phi_A(\vc{k})&=&f_1\left [\int\frac{d^3\vc{p}}{(2\pi)^3}
\Phi_L(\vc{k}+\vc{p})\Phi_L^*(\vc{p})
-(2\pi)^3\delta(\vc{k})\avg{\Phi_L^2(\vc{x})} \right ], \\
\Phi_B(\vc{k})&=&f_2\int\frac{d^3\vc{p}_1}{(2\pi)^3}
\frac{d^3\vc{p}_2}{(2\pi)^3}
\Phi_L^*(\vc{p}_1)\Phi_L^*(\vc{p}_2)\Phi_L(\vc{p}_1+\vc{P}_2+\vc{k}),
\label{PhiAandPhiB}
\end{eqnarray}
where
\begin{eqnarray}
\avg{\Phi_L^2(\vc{x})}=\int\frac{d^3k}{(2\pi)^3} P_\Phi(\vc{k}),
\end{eqnarray}
Here $P_\Phi(\vc{k})$ is the power spectrum of $\Phi_L(\vc{k})$.  

\subsection{Curvature trispectrum}
\label{Sect:InflationTrispec}

The connected part of the four-point function 
$\avg{\Phi(\vc{k}_1)\Phi(\vc{k}_2)\Phi(\vc{k}_3)\Phi(\vc{k}_4)}$ has 
leading order contributions from terms of the form
$\avg{\Phi_A(\vc{k}_1)\Phi_A(\vc{k}_2)\Phi_L(\vc{k}_3)\Phi_L(\vc{k}_4)}$ and
$\avg{\Phi_B(\vc{k}_1)\Phi_L(\vc{k}_2)\Phi_L(\vc{k}_3)\Phi_L(\vc{k}_4)}$.  To
compute the curvature trispectrum, we note that the symmetries with 
respect to exchange of $\Phi(\vc{k}_i)$ and $\Phi(\vc{k}_j)$ in the $\Phi$
four-point function are identical to the symmetries with respect to exchange of
fields in the CMB trispectrum.  We therefore follow the same decomposition
process by first defining 
\begin{eqnarray}
\avg{\Phi(\vc{k}_1)\Phi(\vc{k}_2)\Phi(\vc{k}_3)\Phi(\vc{k}_4)}_c
&=&(2\pi)^3\int d^3K \delta(\vc{k}_1+\vc{k}_2+\vc{K})
\delta(\vc{k}_3+\vc{k}_4-\vc{K})
T_\Phi(\vc{k}_1,\vc{k}_2;\vc{k}_3,\vc{k}_4;\vc{K}),
\end{eqnarray}
and then constructing 
\begin{subequations}
\label{PhiTrispecReduce}
\begin{eqnarray}
T_\Phi(\vc{k}_1,\vc{k}_2;\vc{k}_3,\vc{k}_4;\vc{K})&=&
P_\Phi(\vc{k}_1,\vc{k}_2;\vc{k}_3,\vc{k}_4;\vc{K})
+\int d^3\vc{K}'\left [\delta(\vc{k}_3-\vc{k}_2-\vc{K}+\vc{K}')
P_\Phi(\vc{k}_1,\vc{k}_3;\vc{k}_2,\vc{k}_4;\vc{K}')\right.\nonumber \\
&+&\left.\delta(\vc{k}_4-\vc{k}_2-\vc{K}+\vc{K}')
P_\Phi(\vc{k}_1,\vc{k}_4;\vc{k}_3,\vc{k}_2;\vc{K}')\right ],
\end{eqnarray}
with $P_\Phi$ constructed out of a reduced trispectrum 
$\mathcal{T}_\Phi$ according to 
\begin{eqnarray}
P_\Phi(\vc{k}_1,\vc{k}_2;\vc{k}_3,\vc{k}_4;\vc{K})
&=&\mathcal{T}_\Phi(\vc{k}_1,\vc{k}_2;\vc{k}_3,\vc{k}_4;\vc{K})
+\mathcal{T}_\Phi(\vc{k}_2,\vc{k}_1;\vc{k}_3,\vc{k}_4;\vc{K})\nonumber \\
&+&\mathcal{T}_\Phi(\vc{k}_1,\vc{k}_2;\vc{k}_4,\vc{k}_3;\vc{K})
+\mathcal{T}_\Phi(\vc{k}_2,\vc{k}_1;\vc{k}_4,\vc{k}_3;\vc{K}).
\end{eqnarray}
\end{subequations}
The leading order contributions to the reduced trispectrum are
\begin{subequations}
\label{TPhiContribs}
\begin{eqnarray}
\mathcal{T}_{\Phi_A}(\vc{k}_1,\vc{k}_2;\vc{k}_3,\vc{k}_4;\vc{K})
&=&4f_1^2P_\Phi(K)P_\Phi(k_1)P_\Phi(k_3),\\
\mathcal{T}_{\Phi_B}(\vc{k}_1,\vc{k}_2;\vc{k}_3,\vc{k}_4;\vc{K})
&=&f_2\left [ P_\Phi(k_2)P_\Phi(k_3)P_\Phi(k_4)+
P_\Phi(k_1)P_\Phi(k_2)P_\Phi(k_4)\right ].
\end{eqnarray}
\end{subequations}
The curvature trispectrum induces an angular trispectrum onto
the CMB fluctuations as we shall now see.

\subsection{Angular trispectra}
\label{Sect:TPInfl}

In the linear regime, curvature perturbations generate 
CMB fluctuations as
\begin{eqnarray}
a_l^m &=& 4\pi(-i)^l\int\frac{d^3\vc{k}}{(2\pi)^3}\Phi(\vc{k})g_{al}(k)
Y_l^{m*}(\vh{k}),
\label{TEMoment}
\end{eqnarray}
where $a$ may be the temperature or E-mode multipole moment, $\Phi(\vc{k})$
is the primordial curvature perturbation, and $g_{al}(k)$ denotes the 
radiation transfer function for $a=\Theta,E$.  The multipole moments
$a_l^m$ inherit their statistical properties from $\Phi(\vc{k})$, so that 
in our case, the trispectrum is related directly to integrals of the 
four-point correlation function of $\Phi(\vc{k})$.  

From expression (\ref{TEMoment}), the harmonic four-point function is
related to the $\Phi$ trispectrum by
\begin{eqnarray}
\avg{a_{l_1}^{m_1}b_{l_2}^{m_2}c_{l_3}^{m_3}d_{l_4}^{m_4}}
&=&(4\pi)^4(-i)^{\sum l_i}\int\frac{d^3\vc{k}_1}{(2\pi)^3}\ldots
\frac{d^3\vc{k}_4}{(2\pi)^3}\int d^3K Y_{l_1}^{m_1*}(\vh{k}_1)
Y_{l_2}^{m_2*}(\vh{k}_2)Y_{l_3}^{m_3*}(\vh{k}_3)Y_{l_4}^{m_4*}(\vh{k}_4)
\nonumber \\
&\times &(2\pi)^3g_{al_1}(k_1)g_{bl_2}(k_2)g_{cl_3}(k_3)g_{dl_4}(k_4)
T_\Phi(\vc{k}_1,\vc{k}_2;\vc{k}_3,\vc{k}_4;\vc{K}).
\label{HarmonicToPhiTrispec}
\end{eqnarray}
The reduced trispectrum $\mathcal{T}^{a_{l_1}b_{l_2}}_{c_{l_3}d_{l_4}}(L)$
is then obtained from the reduced $\Phi$ trispectrum simply by replacing
$T_\Phi$ in the above relation and performing the integrals over 
directions $\vh{k}_i$ and $\vh{K}$, so that 
\begin{eqnarray} 
\mathcal{T}^{a_{l_1}b_{l_2}}_{c_{l_3}d_{l_4}}(L)
&=&\left (\frac{2}{\pi}\right )^5\int r_1^2dr_1 r_2^2dr_2
\left (k_1^2 dk_1\right )\cdots \left (k_4^2 dk_4\right )
K^2dKj_L(Kr_1)j_L(Kr_2)
\nonumber \\
&\times &\left [g_{al_1}(k_1)j_{l_1}(k_1r_1)\right ]
\left [g_{bl_2}(k_2)j_{l_2}(k_2r_1)\right ]
\left [g_{cl_3}(k_3)j_{l_3}(k_3r_2)\right ]
\left [g_{dl_4}(k_4)j_{l_4}(k_4r_2)\right ]\nonumber \\
&\times &\mathcal{T}_\Phi(k_1,k_2;k_3,k_4;K)h_{l_1Ll_2}h_{l_3Ll_4},
\end{eqnarray}
where
\begin{eqnarray}
h_{l_1Ll_2}&=&\sqrt{\frac{(2l_1+1)(2l_2+1)(2L+1)}{4\pi}}
\ThreeJ{l_1}{l_2}{L}{0}{0}{0},
\end{eqnarray}
and $j_l(x)$ are the spherical Bessel functions.

Substituting expressions (\ref{TPhiContribs}) into the above, we find that
the reduced trispectrum is given by
\begin{subequations}
\label{PhiInducedTau}
\begin{eqnarray}
\mathcal{T}^{a_{l_1}b_{l_2}}_{c_{l_3}d_{l_4}}(L)&\equiv&
\mathcal{T}_A{}^{a_{l_1}b_{l_2}}_{c_{l_3}d_{l_4}}(L)
+\mathcal{T}_B{}^{a_{l_1}b_{l_2}}_{c_{l_3}d_{l_4}}(L),
\end{eqnarray}
with
\begin{eqnarray}
\mathcal{T}_A{}^{a_{l_1}b_{l_2}}_{c_{l_3}d_{l_4}}(L)
&=&\int r_1^2dr_1 r_2^2dr_2 F_L(r_1,r_2)\alpha^a_{l_1}(r_1)
\beta^b_{l_2}(r_1)\alpha^c_{l_3}(r_2)\beta^d_{l_4}(r_2)
h_{l_1Ll_2}h_{l_3Ll_4},\\
\mathcal{T}_B{}^{a_{l_1}b_{l_2}}_{c_{l_3}d_{l_4}}(L)
&=&\int r^2 dr \beta^b_{l_2}(r)\beta^d_{l_4}(r)
\left [\mu^a_{l_1}(r)\beta^c_{l_3}(r)
+\beta^a_{l_1}(r)\mu^c_{l_3}(r) \right ]h_{l_1Ll_2}h_{l_3Ll_4},
\end{eqnarray}
\end{subequations}
and
\begin{subequations}
\label{PhiInducedDefs}
\begin{eqnarray}
F_L(r_1,r_2)&=&\frac{2}{\pi}\int K^2 dK P_\Phi(K)j_L(Kr_1)j_L(Kr_2),\\
\alpha^a_l(r)&=&\frac{2}{\pi}\int k^2 dk (2f_1)g_{al}(k)j_l(kr),\\
\beta^a_l(r)&=&\frac{2}{\pi}\int k^2 dk P_\Phi(k)g_{al}(k)j_l(kr),\\
\mu^a_l(r)&=&\frac{2}{\pi}\int k^2 dk f_2 g_{al}(k)j_l(kr).
\end{eqnarray}
\end{subequations}
The trispectrum is formed using Eqs. (\ref{TrispecEnforced1}) and 
(\ref{CurlyTFunction}).

To properly evaluate the trispectra, one must extract the
radiation transfer function $g_l(k)$ numerically from an Einstein-Boltzmann
solver as has been done for the temperature bispectrum \cite{Komatsu1}.   
This process is numerically cumbersome and we instead seek an
analytic approximation of its effects. 

\begin{figure}
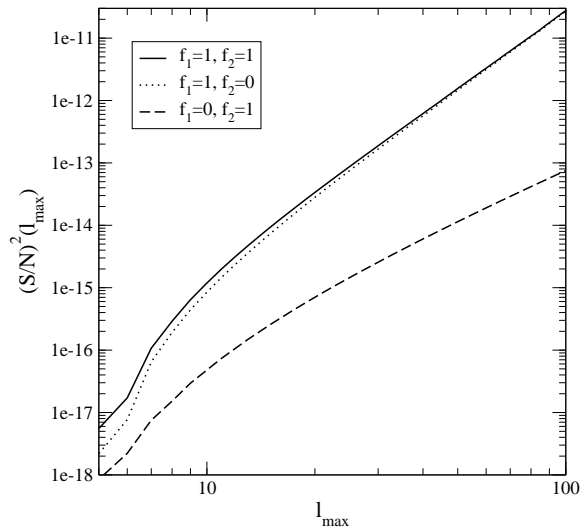

\myfigure{fig2.eps}
\caption{Signal-to-noise ratio in the temperature trispectrum as a function of 
the maximum multipole 
$l_{\text{max}}$ in the Sachs-Wolfe approximation 
(Eq.~\ref{ReducedTrispecSW}).  The dotted line corresponds to $f_1=1,f_2=0$,
the dashed line to $f_1=0,f_2=1$, and the solid line to $f_1=1,f_2=1$.}
\label{fig2}
\end{figure}

For small  multipole moments ($l \ll 100$), CMB temperature fluctuations arise 
mainly from the Sachs-Wolfe effect \cite{SacWol67}.
Here the radiation 
transfer function $g_{\Theta l}(k)$ takes on the simple form
\begin{eqnarray}
g_{\Theta l}(k)&=&\frac{1}{3}j_l(kr_*),
\label{GtlSW}
\end{eqnarray}
with $r_* = \eta_0-\eta_{\text{rec}}$ denoting the conformal time elapsed 
between recombination and the present.  In this regime,
$\alpha^\Theta_l(r)$ and $\mu^\Theta_l(r)$ simplify to
\begin{eqnarray}
\alpha^\Theta_l(r)&=&\frac{2f_1}{3r_*^2}\delta(r-r_*),\label{AlphaSW}\\
\mu^\Theta_l(r)&=&\frac{f_2}{3r_*^2}\delta(r-r_*).\label{MuSW}
\end{eqnarray}
Since the temperature power spectrum is given by
\begin{eqnarray}
C_l^{\text{SW}}&=&\frac{2}{9\pi}\int k^2 dk P_\Phi(k)j_l^2(kr_*),
\label{ClSW}
\end{eqnarray}
the other functions can be related to $C_l^{\text{SW}}$ as  
$F_L(r_*,r_*)=9C_l^{\text{SW}}$ and 
$\beta^\Theta_l(r_*)=3C_l^{\text{SW}}$.  The reduced trispectrum can then be
expressed in terms of $C_l^{\text{SW}}$ as
\begin{eqnarray}
\mathcal{T}^{\Theta_{l_1}\Theta_{l_2}}_{\Theta_{l_3}\Theta_{l_4}}(L)
&=& 9C_{l_2}^{\text{SW}}C_{l_4}^{\text{SW}}\left [4f_1^2C_L^{\text{SW}}+f_2\left (
C_{l_1}^{\text{SW}}+C_{l_3}^{\text{SW}}\right ) \right ]h_{l_1Ll_2}h_{l_3Ll_4}.
\label{ReducedTrispecSW}
\end{eqnarray}
Signal-to-noise ratios calculated from expression (\ref{ReducedTrispecSW})
for various choices of $f_1$ and $f_2$ are shown in Fig. (\ref{fig2}).

To estimate the trispectrum for higher multipoles note that for the
$f_1$ term in (\ref{ReducedTrispecSW}),
$C_{l_2}^{\text{SW}}$ and $C_{l_4}^{\text{SW}}$ appear from an integration over radiation 
transfer functions that is very similar in form to that of the true power
spectrum $C_{l}^{\Theta\Theta}$.  
An approximation to the trispectrum induced by $f_1$ then becomes
\begin{eqnarray}
\mathcal{T}^{\Theta_{l_1}\Theta_{l_2}}_{\Theta_{l_3}\Theta_{l_4}}(L)&\approx&
36h_{l_1Ll_2}h_{l_3Ll_4}f_1^2C_L^{\text{SW}}
C_{l_2}^{\Theta\Theta} C_{l_4}^{\Theta\Theta}
\label{BetterEstimate}
\end{eqnarray}
and should be valid to the extent to which the anisotropies result from
slowly varying
local temperature fluctuations on a thin last scattering surface.
While the $f_2$ term does not have this simple form, we take the
$f_1$ piece as representative.

The Newtonian curvature $\Phi(\vc{k})$ also acts as a source for $E$-mode 
polarization, through the anisotropy of Compton scattering which links the
local quadrupoles of temperature fluctuations to local $E$-mode fluctuations.
Although there is no equivalent to the Sachs-Wolfe approximation for the 
$E$-mode radiation transfer function, we can again take the above approximation
for a slowly-varying source to obtain
\begin{eqnarray}
\mathcal{T}_A{}^{E_{l_1}E_{l_2}}_{E_{l_3}E_{l_4}}(L)
&\approx & 36h_{l_1Ll_2}h_{l_3Ll_4}f_1^2C_L^{\text{SW}}
C_{l_2}^{EE}C_{l_4}^{EE},
\label{PhiInducedE}
\end{eqnarray}
where $C_L^{SW}$ is still the Sachs-Wolfe approximation to the temperature
power spectrum.  Thus, for $f_2=0$, the $E$-mode trispectrum should behave similarly
to the temperature trispectrum.  Mixed $\Theta$ and $E$ trispectra would take on
an analogous form.

\subsection{Signal-to-Noise}
\label{Sect:SNInfl} 

\begin{figure}[tbhp]
\myfigure{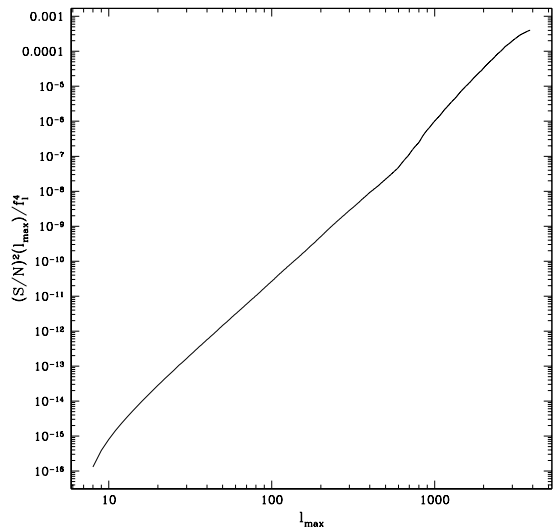}
\caption{ $(S/N)^2/f_1^4$ vs. $l_{\text{max}}$ in the temperature trispectrum, 
using the approximation (\ref{BetterEstimate}).}
\label{fig3}
\end{figure}

We utilize the formalism described in Sec.~\ref{NoiseProperties} to 
calculate the expected signal-to-noise ratio for primordial non-Gaussianity.
Setting $f_2=0$, we use Eq. (\ref{BetterEstimate}) for the temperature 
trispectrum and compute $(S/N)^2$ to the cosmic variance
limit, with Gaussian contamination from gravitational lensing.
Figure~\ref{fig3} shows $(S/N)^2/f_1^4$ as a function of the maximum multipole
moment observed $l_{\text{max}}$.
The tapering of $(S/N)^2$ is due to the fact that the noise contribution from 
lensing becomes significant at small angular scales.  The figure indicates that
the temperature trispectrum may be sensitive to non-Gaussianity for 
$f_1 \alt 10$, although a detailed calculation using expressions 
(\ref{PhiInducedTau}) involving the full radiation transfer function will be 
necessary to place rigorous bounds.  Since polarization trispectra
take a form similar to the temperature, $(S/N)^2$ can be at most 
enhanced by the number of independent trispectra terms.  For $E$ and $\Theta$
combinations this represents a factor of a few.  

Compared with the sensitivity of the temperature bispectrum to primordial
non-Gaussianity, the temperature and polarization trispectra contain 
a comparable amount of information \cite{Komatsu1}. The $(S/N)^2$
in the trispectra can exceed that of the bispectrum if $f_1 \gg 1$ due
to the steep scaling of $f_1^4$.

\section{Discussion\label{Sect:Conclusion}}

We have introduced a complete formalism for the study of 4-point 
correlations in the CMB temperature and polarization fields.  This formalism
should be useful in future tests of the non-Gaussianity of the CMB 
induced in the early universe and by the evolution of structure. It is
also of use in determining the non-Gaussian contributions to errors
in temperature-polarization power spectra measurements.

We have applied these techniques to a particular form of trispectra
motivated by inflation, generalizing previous treatments to higher order
in the initial nonlinearity of the curvature fluctuations. 
Typical slow-roll inflationary models predict
an amplitude to the trispectra that is far from observable and so
a detection of this type of non-Gaussianity would rule out a large class
of models.  We have shown that because of the large number of
trispectra configurations, the sensitivity to initial non-Gaussianity in the
trispectra approach that of the well-studied temperature bispectrum
at high multipoles.  

Trispectra from secondary anisotropies such as gravitational lensing
\cite{HuTrispec}  are expected to be substantially larger and should be
fruitful ground for future studies.   While measurement of these non-Gaussian
signatures will no doubt prove challenging due to foregrounds, systematic
effects and computational cost, the wealth of information potentially
contained therein may justify the large effort that will be required.

\begin{acknowledgments}
We thank E.~Komatsu for useful discussions.  This work was supported
by NASA Grant No. NAG5-10840 and the DOE OJI program. 
\end{acknowledgments}
\appendix
\section{Spin-Weighted Spherical Harmonics\label{Appendix:SpinS}}

We summarize the conventions and properties related to the spin-weighted
spherical harmonics \cite{Goldberg1}.  
A function ${}_sf(\narg{\theta}{\phi})$ on the sphere carries a spin weight $s$ if, 
under a right-handed rotation of the basis $(\vh{e}_\theta,\vh{e}_\phi)$ by an 
angle $\psi$, it transforms according to 
${}_sf(\narg{\theta}{\phi})\rightarrow e^{-is\psi}{}_sf(\narg{\theta}{\phi})$.
For such functions, there exist a complete and orthonormal basis with spin
weight $s$, called the spin-weighted spherical harmonics.
These spin-$s$ spherical harmonics ${}_sY_l^m$ can be constructed from the
ordinary spherical harmonics by application of the raising and lowering 
operators \cite{Goldberg1}.  
Alternately, they are given in terms of rotation matrices
as
\begin{eqnarray}
{}_sY_l^m(\narg{\beta\alpha})
&=& (-1)^m \sqrt{\frac{2l+1}{4\pi}}
D^l_{\ind{-m}{s}}(\rarg{\alpha}{\beta}{\gamma})
e^{is\gamma}.
\label{AppEqn:RotToYlm}
\end{eqnarray}
The Euler angles specify a rotation around the coordinate $\hat{z}$
axis by $\gamma$, followed by a rotation by $\beta$ about $\hat{y}$, then a
rotation by $\alpha$ about the (original) $\hat{z}$ axis.
The rotation matrix is given explicitly by (see, e.g.,~\cite{Varshalovich})
\begin{eqnarray}
D^l_{\ind{-m}{s}}(\rarg{\alpha}{\beta}{\gamma})&=&
e^{-is\alpha}e^{im\gamma}
\left [\frac{(l+m)!(l-m)!}{(l+s)!(l-s)!}\right ]^{\frac{1}{2}}
\sin^{2l}(\beta/2)\nonumber \\
&\times&\sum_k\left (\begin{array}{c}l-s\\k\end{array} \right )
\left (\begin{array}{c}l+s\\k+s-m\end{array} \right )
(-1)^{k+l+s}\cot^{2k+s-m}(\beta/2).
\end{eqnarray}
Note that this convention for $_{s}Y_l^m$ differs from that presented 
in~\cite{Goldberg1,HuAndWhite1} by $(-1)^m$, but corresponds to the Condon-Shortley 
convention for the ordinary spherical harmonics when $s=0$ \cite{ZandS1}.  
Below and throughout the paper we use the shorthand convention for
the arguments of the spin-spherical harmonics and rotation matrices
$\vh{n} = (\narg{\theta}{\phi})$,
$\vh{R} = (\narg{\alpha}{\beta}{\gamma})$
and their differential elements
$d\vh{n} = d\phi\, d\cos\theta$, $d\vh{R} = d\alpha\,  d\cos\beta\,  d\gamma$.

The properties of the spin-weighted spherical harmonics follow from 
those of the rotation matrices.  In the text, we utilize four such properties:
orthogonality, completeness, Clebsch-Gordan expansion, and angle addition.

{\noindent \it Orthogonality:}
\begin{eqnarray}
\int d\vh{R} \,
D^{l*}_{\ind{m}{s}}(\vh{R}) D^{l'}_{\ind{m'}{s'}}(\vh{R}) =  \frac{8\pi^2}{2 l+1} \delta_{ll'} \delta_{mm'}
	\delta_{ss'}
\label{AppEqn:DOrtho}
\end{eqnarray}
implies
\begin{equation}
\int d \vh{n} \,
{}_sY_{l}^{m*}(\vh{n})
{}_sY_{l'}^{m'}(\vh{n})=\delta_{ll'}\delta_{mm'}.
\label{AppEqn:Ortho}
\end{equation}

{\noindent \it Completeness:}
\begin{eqnarray}
\sum_{l m s} D^{l*}_{\ind{m}{s}}(\rarg{\alpha}{\beta}{\gamma}) 
	     D^{l}_{\ind{m}{s}} (\rarg{\alpha'}{\beta'}{\gamma'})
&=& \frac{8\pi^2}{2l+1} 
\delta(\alpha-\alpha')
\delta(\cos\beta-\cos\beta')
\delta(\gamma - \gamma')
\end{eqnarray}
implies
\begin{eqnarray}
\sum_{lm}{}_sY_l^{m*}(\narg{\theta}{\phi}){}_sY_l^m(\narg{\theta'}{\phi'})&=&
\delta(\phi-\phi')\delta(\cos\theta-\cos\theta')\label{AppEqn:Comp}.
\end{eqnarray}

{\noindent \it Clebsch-Gordon relation:}
\begin{eqnarray}
D^{l_1}_{\ind{m_1}{m'_1}}(\vh{R})D^{l_2}_{\ind{m_2}{m'_2}}(\vh{R}) &=& \sum_{LMM'}
(2L+1)
\ThreeJ{l_1}{l_2}{L}{m_1}{m_2}{-M}
\ThreeJ{l_1}{l_2}{L}{m'_1}{m'_2}{-M'}
(-1)^{M+M'}D^L_{\ind{M}{M'}}(\vh{R}),
\label{DFunctionMult}
\end{eqnarray}
implies
\begin{eqnarray}
{}_{s_1} Y_{l_1}^{m_1}(\vh{n})
{}_{s_2} Y_{l_2}^{m_2}(\vh{n})
=
\sqrt{(2l_1+1)(2l_2+1)} \sum_{LMS}(-1)^{M+S} \sqrt{ \frac{2L+1}{4\pi} }
\ThreeJ{l_1}{l_2}{L}{-m_1}{-m_2}{M}
\ThreeJ{l_1}{l_2}{L}{s_1}{s_2}{-S}
{}_S Y_{L}^{M}(\vh{n})\,.
\end{eqnarray}
{\noindent \it Auxiliary {\rm (orthogonality-Clebsch-Gordon)} relation:}
\begin{eqnarray}
\int d \vh{R} \,
D^{l_1}_{\ind{m_1}{m_1'}}(\vh{R})D^{l_2}_{\ind{m_2}{m_2'}}(\vh{R})
D^{l_3}_{\ind{m_3}{m_3'}}(\vh{R})
= 8\pi^2 \ThreeJ{l_1}{l_2}{l_3}{m_1}{m_2}{m_3}\ThreeJ{l_1}{l_2}{l_3}{m_1'}{m_2'}{m_3'},
\label{AppEqn:ThreeDInt}
\end{eqnarray}
implies 
\begin{eqnarray}
\int d\vh{n} {}_{s_1}Y_{l_1}^{m_1}(\vh{n})
{}_{s_2}Y_{l_2}^{m_2}(\vh{n}){}_{s_3}Y_{l_3}^{m_3}(\vh{n})
&=&
\frac{(-1)^{m_1+s_1}}{\sqrt{4\pi}}\left[\prod_{i=1}^3 {2l_i+1}\right]^{1/2}
\ThreeJ{l_1}{l_2}{l_3}{-s_1}{-s_2}{-s_3}\ThreeJ{l_1}{l_2}{l_3}{m_1}{m_2}{m_3}\,
\label{AppEqn:ThreeYlm}
\end{eqnarray}
if $s_1+s_2+s_3=0$.
{\noindent \it Addition theorem:}
\begin{eqnarray}
D^{l*}_{\ind{s_2}{s_1}}(\rarg{\gamma}{\beta}{-\alpha})&=&\sum_m 
D^{l*}_{\ind{-m}{s_1}}(\rarg{\phi'}{\theta'}{0}) D^l_{\ind{-m}{s_2}}
	(\rarg{\phi}{\theta}{0}),
\label{AppEqn:SumStep2}
\end{eqnarray}
implies
\begin{eqnarray}
\sum_m {}_{s_1}Y_l^{m*}(\narg{\theta'}{\phi'}){}_{s_2}Y_l^m(\narg{\theta}{\phi})
&=&\sqrt{\frac{2l+1}{4\pi}}(-1)^{s_2}
{}_{s_2}Y_l^{-s_1}(\narg{\beta}{\alpha})e^{is_2\gamma}.\label{AppEqn:YlmAddition}
\end{eqnarray}
The relationship between the angles (shown in figure~\ref{fig3}) is given 
explicitly by
\begin{eqnarray}
\cot\alpha&=&-\cos\theta'\cot(\phi'-\phi)+\cot\theta
\frac{\sin\theta'}{\sin(\phi'-\phi)},\nonumber \\
\cos\beta&=&\cos\theta\cos\theta'+\sin\theta\sin\theta'\cos(\phi'-\phi),
\nonumber \\
\cot\gamma&=&\cos\theta\cot(\phi'-\phi)
-\cot\theta\frac{\sin\theta'}{\sin(\phi'-\phi)}.
\label{AppEqn:EulerAngles}
\end{eqnarray}
%
%
%
%
\begin{figure}[tbhp]
\myfigure{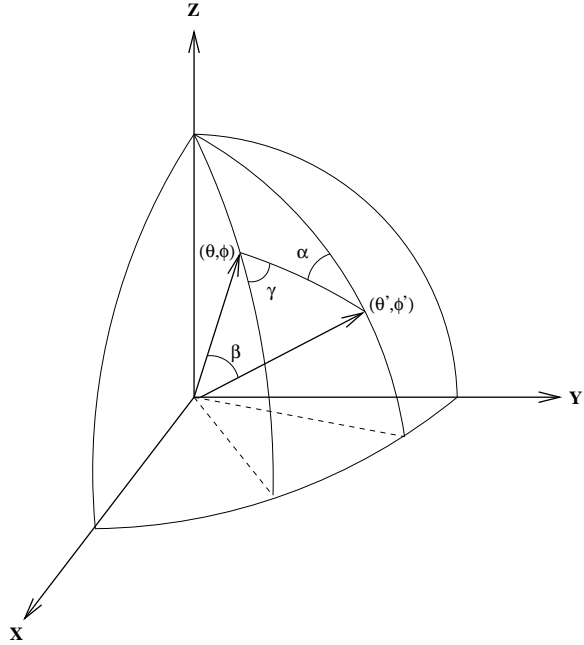}
\caption{Relation between Euler angles $(\rarg{\alpha}{\beta}{\gamma})$ and the original rotation angles $(\narg{\theta}{\phi})$ and $(\narg{\theta'}{\phi'})$  for the weighted sky maps (\ref{WeightedSkyMaps}), with the identification 
$\vh{n}\rightarrow (\narg{\theta'}{\phi'})$, $\vh{q}\rightarrow (\narg{\theta}{\phi})$, so that 
$\phi_{\vh{n}}=\alpha$, and $\vh{n}\cdot\vh{q}=\beta$.}
\label{fig4}
\end{figure}
The addition relation corrects a sign error in 
\cite{HuAndWhite1} and agrees with \cite{NgandLiu}, once one accounts for the differences in the
phase convention for ${}_sY_l^m$.

{\section{Additional Trispectra  Properties}
\label{Appendix:Additional}

\subsection{Symmetries}
\label{Appendix:FixedFields}}

Here, we present the symmetry properties of trispectra when only the angular
momentum labels are permuted, keeping the field labels fixed.  Such a
representation is redundant but can be useful if the diagonal length $L$ is 
related to some physical quantity, as is true for the inflationary trispectra
[see Eq.~(\ref{PhiInducedTau})].

Using the three CMB fields 
$\{\Theta,E,B\}$, 15 distinct four-point functions can be constructed, 
with the field contents
\begin{eqnarray}
 \xf\xf\xf\xf&\in&\{\Theta\Theta\Theta\Theta, EEEE,BBBB\},\nonumber \\
 \xf\xf\xf\yf&\in&\{\Theta\Theta\Theta E,\Theta\Theta\Theta B,EEE\Theta,
EEEB,BBB\Theta,BBBE \},\nonumber \\
\xf\xf\yf\yf&\in&\{\Theta\Theta EE,\Theta\Theta BB,EEBB\},\text{ and}\nonumber \\
\xf\xf\yf\zf&\in&\{\Theta\Theta EB,EE\Theta B,BB\Theta E\}.
\end{eqnarray}
For each case, the use of identical fields results in a restricted set of 
permutation symmetries.

The case for $\avg{\Theta\Theta\Theta\Theta}$ has been worked out in detail in 
\cite{HuTrispec}, where the 
trispectrum was shown to be composed from a reduced form, thus incorporating 
the
$4!=24$ possible permutations of $l_i$'s which leave the four-point harmonic
function unchanged.  Following the treatment in~\cite{HuTrispec}, for 
trispectra of the form $\xf\xf\xf\xf$, permutation symmetry of $l_i$ requires that the trispectrum 
$Q^{\xf_{l_1}\xf_{l_2}}_{\xf_{l_3}\xf_{l_4}}(L)\equiv Q^{l_1l_2}_{l_3l_4}(L)$ obey
the constraints
\begin{subequations}
\label{XXXXConstraints}
\begin{eqnarray}
Q^{l_1l_2}_{l_3l_4}(L)&=&(-1)^{\Sigma_U}Q^{l_2l_1}_{l_3l_4}(L)
=(-1)^{\Sigma_L}Q^{l_1l_2}_{l_4l_3}(L)
=(-1)^{\Sigma_U+\Sigma_L}
Q^{l_3l_4}_{l_1l_2}(L),
\label{XXXXConstraint1}\\
Q^{l_1l_2}_{l_3l_4}(L)&=&\sum_{L'}(-1)^{l_2+l_3}(2L+1)
\SixJ{l_1}{l_2}{L}{l_4}{l_3}{L'}
Q^{l_1l_3}_{l_2l_4}(L'),
\label{XXXXConstraint2} 
\end{eqnarray}
and
\begin{eqnarray}
Q^{l_1l_2}_{l_3l_4}(L)&=&\sum_{L'}(-1)^{L+L'}(2L+1)
\SixJ{l_1}{l_2}{L}{l_3}{l_4}{L'}
Q^{l_1l_4}_{l_3l_2}(L'),\label{XXXXConstraint3}
\end{eqnarray}
\end{subequations}
where $\Sigma_U\equiv L+l_1+l_2$, and $\Sigma_L\equiv L+l_3+l_4$.

For trispectra of the form $\xf\xf\xf\yf$, the permutations $(123)$ on the $l$ indices
are allowed, so that the constraints are given by
\begin{subequations}
\label{XXXYConstraints}
\begin{eqnarray}
Q^{\xf_{l_1}\xf_{l_2}}_{\xf_{l_3}\yf_{l_4}}(L)&=&
(-1)^{\Sigma_U}Q^{\xf_{l_2}\xf_{l_1}}_{\xf_{l_3}\yf_{l_4}}(L)
\label{XXXYConstraint1}\\
&=&\sum_{L'}(-1)^{l_2+l_3}(2L+1)
\SixJ{l_1}{l_2}{L}{l_4}{l_3}{L'}
Q^{\xf_{l_1}\xf_{l_3}}_{\xf_{l_2}\yf_{l_4}}(L')
\label{XXXYConstraint2} \\
&=&\sum_{L'}(-1)^{L+L'}(2L+1)
\SixJ{l_1}{l_2}{L}{l_3}{l_4}{L'}
Q^{\xf_{l_3}\xf_{l_2}}_{\xf_{l_1}\yf_{l_4}}(L'),
\label{XXXYConstraint3}
\end{eqnarray}
\end{subequations}
where the last two relations come from Eq. (\ref{Recoupling}).
These exhaust the allowed $3!=6$ permutation symmetries.

For trispectra related to 
$\avg{\xf_{l_1}^{m_1}\yf_{l_2}^{m_2}\xf_{l_3}^{m_3}\yf_{l_4}^{m_4}}$, the permutations
$(l_1\leftrightarrow l_3)$ and $(l_2\leftrightarrow l_4)$ are separately 
allowed, so that there should be $2\cdot 2=4$ permutations to account for.
In the $\xf\xf\yf\yf$ basis, the permutation symmetries imply that
\begin{eqnarray}
Q^{\xf_{l_1}\xf_{l_2}}_{\yf_{l_3}\yf_{l_4}}(L)=
(-1)^{\Sigma_U}Q^{\xf_{l_2}\xf_{l_1}}_{\yf_{l_3}\yf_{l_4}}(L)
=(-1)^{\Sigma_L}Q^{\xf_{l_1}\xf_{l_2}}_{\yf_{l_4}\yf_{l_3}}(L)
=(-1)^{\Sigma_U+\Sigma_L}Q^{\xf_{l_2}\xf_{l_1}}_{\yf_{l_4}\yf_{l_3}}(L).
\label{XXYYConstraints}
\end{eqnarray}
Expressions in the other bases can be derived through the use of the recoupling
relations (\ref{Recoupling}).  

Lastly, for trispectra related to four-point functions 
$\avg{\xf_{l_1}^{m_1}\xf_{l_2}^{m_2}\yf_{l_3}^{m_3}\zf_{l_4}^{m_4}}$
the only permutation symmetry allowed is the exchange of $l_1$ and $l_2$.
Accordingly, the trispectra obey
\begin{eqnarray}
Q^{\xf_{l_1}\xf_{l_2}}_{\yf_{l_3}\zf_{l_4}}(L)&=&(-1)^{\Sigma_U}
Q^{\xf_{l_2}\xf_{l_1}}_{\yf_{l_3}\zf_{l_4}}(L),
\label{XXYZConstraints}
\end{eqnarray}
where, again, the recoupling relations can be used to find the symmetry 
constraints in the other bases.

The signal-to-noise ratio for trispectra with fixed field configurations can
be obtained from expression (\ref{SNtemp}) by using the recoupling relations
(\ref{Recoupling}) to permute the field symbols into that of the fixed field
representation.  The results can be simplified by using the following 
identities for the 6-j symbols,
\begin{eqnarray}
\sum_e (2e+1)\SixJ{a}{b}{e}{c}{d}{f}\SixJ{a}{b}{e}{c}{d}{g}
=\frac{\delta_{fg}}{2f+1},
\end{eqnarray}
and
\begin{eqnarray}
\sum_e(-1)^{e+f+g}(2e+1)\SixJ{a}{b}{e}{c}{d}{f}\SixJ{a}{b}{e}{d}{c}{g}
=\SixJ{a}{c}{g}{b}{d}{f}.
\end{eqnarray}
Rewriting the signal-to-noise ratio in terms of the fixed field configurations
transfers the redundancies in the field permutations into redundancies in the
$l$ configurations, so that the restrictions on the sum over $l_i$'s typically
become relaxed.

\subsection{Measurement}
\label{Appendix:Measurement}

Direct measurement of the trispectrum using the estimator of
Eq. (\ref{TrispecEstimator}) is computationally expensive due in part 
to the quintuple sum over $m$'s.  Since the $m$ dependence of the four-point function
simply reflects the rotational invariance, it is useful to find an 
estimator that employs these symmetries in a more efficient way.
The following construction parallels that of the temperature bispectrum \cite{Spergel1}
and trispectrum \cite{HuTrispec} and takes into account the subtleties
due to the spin-2 behavior of the polarization fields.
For the temperature and polarization fields, one can define a set of 
weighted sky maps $e_l^\alpha(\vh{q})$, where
\begin{subequations}
\label{WeightedSkyMaps}
\begin{eqnarray}
e_l^\Theta(\vh{q})&=&\sqrt{\frac{2l+1}{4\pi}}\int d\vh{n}\Theta(\vh{n})
P_l(\vh{n}\cdot\vh{q}),\\
e_l^E(\vh{q})&=&\sqrt{\frac{2l+1}{4\pi}\frac{(l-2)!}{(l+2)!}}\int d\vh{n}
P_l^2(\vh{n}\cdot\vh{q})
\left [Q(\vh{n})\cos 2\phi_{\vh{n}}+U(\vh{n})\sin 2\phi_{\vh{n}} \right ],
\end{eqnarray}
and 
\begin{eqnarray}
e_l^B(\vh{q})&=&\sqrt{\frac{2l+1}{4\pi}\frac{(l-2)!}{(l+2)!}}\int d\vh{n}
P_l^2(\vh{n}\cdot\vh{q})
\left [U(\vh{n})\cos 2\phi_{\vh{n}}-Q(\vh{n})\sin 2\phi_{\vh{n}} \right ].
\end{eqnarray}
\end{subequations}
The angle $\phi_{\vh{n}}$ is the angle between the great circles defined
by $(\vh{n},\vh{z})$ and $(\vh{n},\vh{q})$,  and serves to transform the 
Stokes parameters from the spherical polar basis to the great circle basis
(see Fig.~\ref{fig4}).
Using these maps, the quantity 
\begin{eqnarray}
\hat{Q}^{\wf_{l_1}\xf_{l_2}}_{\yf_{l_3}\zf_{l_4}}(L)
&\equiv&
\hat{T}^{\wf_{l_1}\xf_{l_2}}_{\yf_{l_3}\zf_{l_4}}(L)
+\hat{G}^{\wf_{l_1}\xf_{l_2}}_{\yf_{l_3}\zf_{l_4}}(L)
\end{eqnarray}
can be estimated by expanding the harmonic coefficients in 
the direct estimator (\ref{TrispecEstimator}) back into fields,
expanding the Wigner 3-j symbols using equation (\ref{AppEqn:ThreeYlm}), resulting in 
the expression
\begin{eqnarray}
\ThreeJ{l_1}{l_2}{L}{0}{0}{0}\ThreeJ{l_3}{l_4}{L}{0}{0}{0}
\hat{Q}^{\wf_{l_1}\xf_{l_2}}_{\yf_{l_3}\zf_{l_4}}(L)
&=&(2L+1)
\int \frac{d\vh{n}_1}{4\pi}\frac{d\vh{n}_2}{4\pi}P_L(\vh{n}_1\cdot\vh{n}_2)
e_{l_1}^\wf(\vh{n}_1)e_{l_2}^\xf(\vh{n}_1)
e_{l_3}^\yf(\vh{n}_2)e_{l_4}^\zf(\vh{n}_2).
\label{Measurement}
\end{eqnarray}
The 3-j symbols impose the constraint $l_1+l_2+L=$even, so the above expression
does not allow measurement of modes with $l_1+l_2+L=$odd.  It may be computationally
advantageous to compute the double integral in Eq. (\ref{Measurement}) as a 
single sum in multipole 
space by noting that they individually return the harmonic decomposition of a product of
two $e$-fields \cite{KomatsuPhD}.
One must also 
account for complications arising from realistic issues, including
the leakage between the $E$ and $B$ modes due to incomplete sky coverage, for example,
by Monte-Carlo techniques.

\subsection{Flat sky approximation}

A sufficiently small patch of sky ($\theta \ll 1$) can be considered flat. 
In this limit it is computationally
and conceptually advantageous 
to consider the Fourier representation of the trispectrum.  Here
we establish the relationship between the angular and flat-sky trispectra.

In the flat-sky approximation, the temperature 
and polarization fields are expanded in Fourier modes as
\begin{subequations}
\label{FlatSkyExpansions}
\begin{eqnarray}
\Theta(\vh{n})&=&\int\frac{d^2\vc{l}}{(2\pi)^2}\Theta(\vc{l})
e^{i\vc{l}\cdot\vh{n}},\label{FlatSkyThetaExp}\\
{}_\pm A(\vh{n})&=&-\int\frac{d^2\vc{l}}{(2\pi)^2}{}_\pm A(\vc{l})
e^{\pm 2i(\phi_{\vc{l}}-\phi)}e^{i\vc{l}\cdot\vh{n}},
\label{FlatSkyQU}
\end{eqnarray}
\end{subequations}
where $\phi_{\vc{l}}$ is the azimuthal angle of $\vc{l}$, and the Stokes 
parameters are defined in a spherical basis.  Again, the $E$ and $B$ modes are 
defined as 
\begin{eqnarray}
{}_\pm A(\vc{l})&=&E(\vc{l})\pm iB(\vc{l}),
\end{eqnarray}
and the two-point angular correlation functions reduce to
\begin{eqnarray}
\avg{\xf^*(\vc{l})\xf'(\vc{l}')}&=&(2\pi)^2\delta(\vc{l}-\vc{l}')C_{(l)}^{\xf\xf'}.
\label{FlatSkyPowerSpec}
\end{eqnarray}
The connected part of the four-point correlation function can be written as
\begin{eqnarray}
\avg{\wf(\vc{l}_1)\xf(\vc{l}_2)\yf(\vc{l}_3)\zf(\vc{l}_4)}_c&=&(2\pi)^2
\delta(\vc{l}_1+\vc{l}_2+\vc{l}_3+\vc{l}_4)
T^{(\wf_{l_1}\xf_{l_2})}_{(\yf_{l_3}\zf_{l_4})}(l_{12},l_{13}),
\label{FlatSkyFourPoint}
\end{eqnarray}
where $l_{12}$ and $l_{13}$ denote the lengths of the two diagonals.  This
can be broken up into pieces corresponding to distinct pairings, so that
\begin{eqnarray}
T^{(\wf_{l_1}\xf_{l_2})}_{(\yf_{l_3}\zf_{l_4})}(l_{12},l_{13})&=&
P^{(\wf_{l_1}\xf_{l_2})}_{(\yf_{l_3}\zf_{l_4})}(l_{12})+
P^{(\wf_{l_1}\xf_{l_3})}_{(\yf_{l_2}\zf_{l_4})}(l_{13})+
P^{(\wf_{l_1}\xf_{l_4})}_{(\yf_{l_3}\zf_{l_2})}(l_{14}),
\label{FlatSkyTDecomposition}
\end{eqnarray}
with $l_{14}$ being a function of $l_{12}$ and $l_{13}$.  The second and third
terms can be projected onto the first pairing.  Denoting the trispectrum with
the projected terms as $T^{(\wf_{l_1}\xf_{l_2})}_{(\yf_{l_3}\zf_{l_4})}(L)$,
the four-point function becomes
\begin{eqnarray}
\avg{\wf(\vc{l}_1)\xf(\vc{l}_2)\yf(\vc{l}_3)\zf(\vc{l}_4)}_c&=&(2\pi)^2
\int d^2\vc{L}
\delta(\vc{l}_1+\vc{l}_2+\vc{L})
\delta(\vc{l}_3+\vc{l}_4-\vc{L})
T^{(\wf_{l_1}\xf_{l_2})}_{(\yf_{l_3}\zf_{l_4})}(L),
\label{FlatSkyTrispec}
\end{eqnarray}
where we used a decomposition of the delta function
\begin{eqnarray}
\delta(\vc{l}_1+\vc{l}_2+\vc{l}_3+\vc{l}_4)
&=&\int d^2\vc{L}\delta(\vc{l}_1+\vc{l}_2+\vc{L})
\delta(\vc{l}_3+\vc{l}_4-\vc{L}).
\label{FlatSkyDeltaFunction}
\end{eqnarray}
Separating the pairings according to the prescription in Eq. 
(\ref{FlatSkyTDecomposition}), the flat-sky four-point function is separated
into
\begin{eqnarray}
\avg{\wf(\vc{l}_1)\xf(\vc{l}_2)\yf(\vc{l}_3)\zf(\vc{l}_4)}_c&=&(2\pi)^2
\int d^2\vc{L}\left \{\delta(\vc{l}_1+\vc{l}_2+\vc{L})
\delta(\vc{l}_3+\vc{l}_4-\vc{L})
P^{(\wf_{l_1}\xf_{l_2})}_{(\yf_{l_3}\zf_{l_4})}(L)\right.\nonumber \\
&+&\delta(\vc{l}_1+\vc{l}_3+\vc{L})
\delta(\vc{l}_2+\vc{l}_4-\vc{L})
P^{(\wf_{l_1}\yf_{l_3})}_{(\xf_{l_2}\zf_{l_4})}(L)\nonumber \\
&+&\left.\delta(\vc{l}_1+\vc{l}_4+\vc{L})
\delta(\vc{l}_3+\vc{l}_2-\vc{L})
P^{(\wf_{l_1}\zf_{l_4})}_{(\yf_{l_3}\xf_{l_2})}(L)\right \}.
\label{FlatSkyP}
\end{eqnarray}

To relate the above expression for the four-point function in the flat sky
approximation to the full-sky expression, we use the relation between the
Fourier coefficients $\xf(\vc{l})$ and $\xf_l^m$,
\begin{subequations}
\label{XlXlmRelation}
\begin{eqnarray}
\xf(\vc{l})&=&\sqrt{\frac{4\pi}{2l+1}}\sum_m i^m\xf_l^me^{im\phi_l}
\label{XltoXlm}
\end{eqnarray}
and the inverse
\begin{eqnarray}
\xf_l^m&=&\sqrt{\frac{2l+1}{4\pi}}i^{-m}\int\frac{d\phi_l}{2\pi}
e^{-im\phi_l}\xf(\vc{l}),
\end{eqnarray}
\end{subequations}
where $\xf\in\{\Theta,E,B\}$.  The phase convention chosen here differs from 
those in \cite{White1,Hu2}, due to differences in the choice of phase in
the definition of the spin-weighted spherical harmonics.

The plane waves in the $\delta$ functions are further decomposed into 
spherical harmonics using the relation
\begin{eqnarray}
e^{i\vc{l}\cdot\vh{n}}&\approx&\sqrt{\frac{2\pi}{l}}\sum_m (-i)^mY_l^m 
e^{-im\phi_l},
\label{PlaneWaveToYlm}
\end{eqnarray}
where the approximation is valid for small angles (or large multipoles $l$).  

If the four-point function has an even net parity, then 
$P^{\wf_{l_1}\xf_{l_2}}_{\yf_{l_3}\zf_{l_4}}(L)$ is independent of the orientation of
the quadrilaterals, so that the integrals over the azimuthal angles 
$\phi_{\vc{l}}$ can be performed, and we obtain the desired relation
\begin{eqnarray}
P^{\wf_{l_1}\xf_{l_2}}_{\yf_{l_3}\zf_{l_4}}(L)&\approx& 
\frac{2L+1}{4\pi}\sqrt{(2l_1+1)\ldots(2l_4+1)}
\ThreeJ{l_1}{l_2}{L}{0}{0}{0}\ThreeJ{l_3}{l_4}{L}{0}{0}{0}
P^{(\wf_{l_1}\xf_{l_2})}_{(\yf_{l_3}\zf_{l_4})}(L).
\label{FlatToFullTrispec}
\end{eqnarray}
For odd net parity, the Wigner 3-j symbols should be reinterpreted as
their analytic continuation (see \cite{Hu2}, Eq. (B1) and Appendix C3).

\end{document}